\documentclass[useAMS,usegraphicx,usenatbib]{mn2e}
\usepackage{amsmath}
\newcommand{\ppmap}{{\tt PPMAP }}

\title[Temperature as a third dimension in column-density mapping of dusty astrophysical structures]{Temperature as a third dimension in column-density mapping of dusty astrophysical structures associated with star formation}
\author[K. A. Marsh et al.]{K. A. Marsh\thanks{E-mail:
Ken.Marsh@astro.cf.ac.uk}, A. P. Whitworth, \& O. Lomax \\
School of Physics and Astronomy, Cardiff University, Cardiff CF24 3AA, UK}
\begin{document}

\pagerange{\pageref{firstpage}--\pageref{lastpage}} \pubyear{2002}

\maketitle

\label{firstpage}

\begin{abstract}
We present {\tt PPMAP}, a Bayesian procedure that uses
images of dust continuum emission at multiple wavelengths to produce 
resolution-enhanced image cubes of differential column-density as a function of
dust temperature and position. \ppmap  is based on the generic ``point process" 
formalism, whereby the system of interest (in this case, a dusty astrophysical
structure such as a filament or prestellar core) is represented by a 
collection of points in a suitably defined state space. 
It can be applied to a variety of observational data, such as
{\it Herschel\/} images, provided only that the image intensity is delivered 
by optically thin dust in thermal equilibrium. \ppmap takes full account of the 
instrumental point spread functions and does not require all images 
to be degraded to the same resolution. We present the results of testing
using simulated data for a prestellar core and a fractal
turbulent cloud, and demonstrate its performance with real data
from the Hi-GAL survey. Specifically, we analyse observations
of a large filamentary structure in the CMa OB1 giant molecular cloud.
Histograms of differential column-density indicate that the warm material
($T\stackrel{>}{_\sim}13$ K) is distributed log-normally, consistent with
turbulence, but the column-densities of the cooler material are distributed
as a high density tail, consistent with the effects of self-gravity.
The results illustrate
the potential of \ppmap to aid in distinguishing between different physical 
components along the line of sight in star-forming clouds, and aid
the interpretation of the associated PDFs of column density.
\end{abstract}

\begin{keywords}
techniques: high angular resolution --- techniques: image processing --- methods: data analysis --- stars: formation --- submillimetre: ISM --- ISM: clouds.
\end{keywords}

\section{Introduction}

Observations of thermal dust emission from Galactic structures, such as
starless cores, filaments, and bubbles in the interstellar medium (ISM),
can provide key information on the initial conditions for star formation.
Significant advances in the modelling of these structures have been 
made possible by the
availability of submillimetre imaging data at multiple wavelengths, such as
that provided by {\it Herschel\/} \citep{pilb10} and various ground-based 
telescopes. The data carry information on physical parameters such as
the density and temperature structure of prestellar
cores and their filamentary environments, as well as that of the
turbulent medium from which those structures are believed to evolve. 
The parameter estimation typically involves the fitting of modified blackbody 
models to the observed spectral energy distributions (SEDs).
This may carried out on a pixel-by-pixel basis so as to produce
maps of integrated column density and mean line-of-sight dust temperature
(sometimes referred to as {\it column\/} temperature).

In the standard procedure
\citep[see, for example,][]{kon10,per10,bern10},
the images at all wavelengths are first smoothed to a common spatial resolution
which, in the case of {\it Herschel\/} data, means the resolution at
the longest wavelength, i.e., 500 $\mu$m. A variant of this technique
\citep{palm13} uses spatial filtering to restore (with the possible exception
of the cooler structures) the 250 $\mu$m spatial resolution.
Three physical assumptions underlying these procedures are:
(i) the dust along a given line of sight has 
uniform temperature, (ii) the ratio of gas to dust is uniform, and
(iii) the dust opacity law is constant and represented by a power-law,
$\kappa_\lambda\propto\lambda^{-\beta}$, as a function of wavelength,
$\lambda$, with $\beta$ normally being taken as 2. The latter assumption 
may be relaxed by allowing $\beta$ to vary \citep[e.g.,][]{gord14},
although care must be exercised in order to avoid spurious correlations
due to the degeneracy between dust temperature, $T$, and $\beta$
\citep{kelly12,ven13}.

Since the structures of interest are, 
in general, not isothermal, assumption (i) is often a poor one. This is 
particularly true of structures such as prestellar cores, which have large 
temperature gradients. Those gradients can have significant effects on
parameter values estimated from SED fits. For example,
\citet{mal11} showed that line-of-sight temperature variations can lead to
underestimates in mass. Also, \citet{shet09} found that the temperature 
variations result in poor fits of the peaks of SEDs to the models, and that
temperature estimates based on simple SED fits can provide only upper
limits to the coldest temperatures along the line-of-sight.

If observational images are available at several (at least three) wavelengths, 
there is information contained in the data
which allows us to constrain the 
distribution of temperature on the line of sight. We accomplish this goal by 
using the set of observed images to produce an image cube consisting of a 
stack of 2D images of differential column-density, where each image in the 
stack represents the column-density at a different dust temperature. This is 
an application of the more generic ``point process" algorithm 
\citep{rich87,rich92}, and we therefore refer to the new procedure as {\it 
point process mapping}, or {\tt PPMAP}.

\section[]{Mathematical Basis}

\subsection[]{Point Processes}

A point process is defined as a random set of points in a suitably-defined 
state space. It provides a conceptual framework for representing an 
astrophysical system as a collection of primitive ``objects'',\footnote{In 
this paper we use the term ``object" to refer to one of the primitive 
building blocks of an astrophysical structure such as a filament or 
core, rather than to the entire structure itself.} each of which is 
characterised by a set of parameters. Those parameters then constitute the 
axes of a ``single-object state space", so that the system itself is 
represented by a distribution of points in such a space.  In the present 
context, the system may be a filament, bubble, core, or molecular cloud, and 
the constituent ``objects" are small building blocks, each of unit 
column-density and uniform temperature. Each such object is then characterised 
by three parameters: 2D position projected onto the plane of the sky $(x,y)$, 
and dust temperature $(T)$; it can thus be represented as a point in a 
three-dimensional state space.  We divide the state space into a rectangular 
grid of $N_{\rm st}$ cells corresponding to
the total number of states; the column-density 
distribution as a function of position and temperature is then defined by the 
set of occupation numbers of those cells, represented by 
vector ${\mathbf \Gamma}$ which is referred to as the ``state" of the system.

\subsection[]{Measurement Model}

We assume that the astrophysical structure is optically thin to the radiation 
emitted by dust at all observed wavelengths. Consequently the images are the 
superposition of the instrumental responses to all of the individual component 
objects, whose number is denoted by $N$.  Each object is defined to have unit 
column-density and a spatial profile corresponding to a circular Gaussian 
whose full width at half maximum corresponds to 2 pixels in the positional 
grid. The measurement model can then be expressed as:
\begin{equation}
{\mathbf d} = {\mathbf A}{\mathbf\Gamma} + {\mathbf\mu}\,.
\label{eq1}
\end{equation}
Here, ${\mathbf d}$ is the measurement vector whose $m^{\rm th}$ component 
represents the pixel value at location $(X_m,Y_m)$ in the observed image at 
wavelength $\lambda_m$; $\;{\mathbf\mu}$ is the measurement noise,\footnote{The
noise term includes background fluctuations if a sky background has been 
subtracted from the observational images.} assumed to be a Gaussian random 
process with covariance ${\mathbf C}_\mu$; $\;{\mathbf A}$ is the system 
response matrix whose $mn^{\rm th}$ element expresses the response of the $m^{\rm th}$ measurement to an object which occupies the $n^{\rm th}$ cell in the state space, corresponding to spatial location $(x_n,y_n)$ and dust temperature $T_n$; it is given by
\begin{equation}
A_{mn} = H\!_{\lambda_m}\!(X_m\!\!-\!\!x_n, Y_m\!\!-\!\!y_n)\,K\!_{\lambda_m}\!(T_n)B\!_{\lambda_m}\!(T_n)\,\kappa(\lambda_m)\,\Delta\Omega_m\,.
\label{eq0}
\end{equation}
Here, $H_\lambda(x,y)$ is the convolution of the point spread function (PSF) 
at wavelength $\lambda$ with the profile of an individual object; 
$K_\lambda(T)$ is a possible colour correction to the model fluxes
due the finite bandwidth of the observations\footnote{Typically specified as 
a lookup table derived using the instrumental pass band shapes.}; 
$B_\lambda(T)$ is the Planck function; 
$\Delta\Omega_m$ is the solid angle subtended by the $m^{\rm th}$ 
pixel and $\kappa(\lambda)$ is the dust opacity 
law. We could use any appropriate functional form for the latter, but
for present purposes we adopt a simple power law with a 
constant index\footnote{Our choice of constant $\beta$ in the present case is 
motivated by the difficulty of constraining the opacity law with the limited 
coverage of long wavelengths in the {\it Herschel\/} data set. A future 
version of the algorithm will incorporate
$\beta$ as a state variable and the observations will be 
supplemented by ground-based 850 $\mu$m data.} of $\beta=2$, i.e.,
\begin{equation}
\kappa(\lambda) = 0.1\,{\rm cm}^2\,{\rm g}^{-1}\,\left(\frac{\lambda}{300\,\mu{\rm m}}\right)^{-2}\,.
\label{eq12}
\end{equation}
Eq. (\ref{eq12}) provides a reasonably good approximation (to within $\sim50$\%)
when applied to observations of starless cores \citep{roy14}. 
The reference opacity (0.1 cm$^2$ g$^{-1}$ at 300 
$\mu$m) is defined with respect to total mass (dust plus gas). Although
observationally determined, it is consistent with a gas to dust ratio of 
100 \citep{hild83}.

The state vector, ${\mathbf\Gamma}$, is regarded as another random 
process; its individual components, $\Gamma_n$, are assumed to be 
statistically independent and binomially distributed {\it a priori\/}, i.e.
\begin{equation}
P(\Gamma_n) = \begin{cases} \frac{N!}{\Gamma\!_n!(N-\Gamma\!_n)!}\,p^{\Gamma\!_n}(1 - p)^{N-\Gamma_n} & \mbox{if } \Gamma_n\in\{0,\ldots,N\}, \\ 0, & \mbox{otherwise. } \end{cases}
\label{eq2}
\end{equation}
Here $p$ is the probability that any given cell is occupied when there is only one object present, i.e. $p = 1/N_{\rm st}$. 
The {\it a priori\/} mean of $\Gamma_n$ is equal to the constant
value $\eta$ for all $n$, where $\eta= N_0/N_{\rm st}$
and $N_0$ is the {\it a priori\/} expectation number of objects.
For sufficiently large $N$ (in practice, $N\stackrel{>}{_\sim}20$), 
the deMoivre-Laplace theorem enables Eq. (\ref{eq2}) to
be approximated by a Gaussian, such that:
\begin{equation}
P(\Gamma_n) = \frac{1}{\sigma\sqrt{2\pi}} \exp 
\frac{-(\Gamma_n - \eta)^2}{2\sigma^2}
\label{eq3a}
\end{equation}
where $\sigma=\sqrt{\eta(1-p)}$.
\smallskip

In either case, the {\it a priori\/} distribution of possible states is 
given by
\begin{equation}
P({\mathbf\Gamma}) = \prod_{n=1}^{N_{\rm st}} P(\Gamma_n).
\label{eq3}
\end{equation}

\subsection[]{Solution Methodology}

The goal of the procedure is to estimate ${\mathbf\Gamma}$ given the 
data, ${\mathbf d}$. The estimation is based on minimising the mean square 
error, so that the optimal estimate is then the {\it a posteriori\/} average 
value of ${\mathbf\Gamma}$, given by:
\begin{equation}
\rho({\mathbf z}_n|{\mathbf d}) \equiv E(\Gamma_n|{\mathbf d}) = \sum_{\mathbf\Gamma} \Gamma_n P({\mathbf\Gamma}|{\mathbf d})\,.
\label{eq4}
\end{equation}
Here, ${\mathbf z}_n$ is a 3-dimensional vector representing the 
coordinates $(x_n,y_n,T_n)$ of the $n^{\rm th}$ cell in state space. The 
conditional probability, $P({\mathbf\Gamma}|{\mathbf d})$, is given by 
Bayes' rule,
\begin{equation}
P({\mathbf\Gamma}|{\mathbf d}) = \frac{P({\mathbf d}|{\mathbf\Gamma})\,P({\mathbf\Gamma})}{P({\mathbf d})}\,,
\label{eq5}
\end{equation}
where $P(\Gamma)$ is given by (\ref{eq3}), $P({\mathbf d})$ serves as a 
normalisation factor, 
\begin{equation}
\ln P({\mathbf d}|{\mathbf\Gamma}) = -\frac{1}{2}({\mathbf d} - {\mathbf A}
{\mathbf\Gamma})^{\rm T} {\mathbf C}_\mu^{-1}({\mathbf d}-{\mathbf A}{\mathbf\Gamma})
+ {\rm const.}
\label{eq6}
\end{equation}
and ${\rm T}$ denotes the transpose.

We refer to $\rho({\mathbf z}|{\mathbf d})$ as a density since it represents 
the average local density of occupied cells in the state space of position 
and temperature. Its estimation is a generic problem in statistical mechanics, 
and its solution has been discussed previously in connection with acoustical 
imaging \citep{rich87}, target tracking \citep{rich92}, and the detection of 
planets using interferometric data \citep{marsh06}. We use a stepwise approach 
in which we start by artificially increasing the measurement noise to the 
point at which the measurements contribute essentially no information; the 
optimal solution is then simply the {\it a priori\/} mean density which is 
flat everywhere. We then gradually decrease the noise back down to the true 
value, updating $\rho({\mathbf z}|{\mathbf d})$ at each step. The process can 
be regarded as a time sequence of noisy measurements whose cumulative effect 
is to build the signal to noise ratio (SNR) back up to the correct value. 
The ``time" corresponds to a progress variable, $t$, representing the 
degree of conditioning on the data, and its value increases from 0 to 1 
during the estimation process. On this basis we rewrite our measurement model 
as
\begin{equation}
{\mathbf d}(t) = {\mathbf A}{\mathbf\Gamma}(t) + {\mathbf\nu}(t)\,,
\label{eq7}
\end{equation}
where ${\mathbf\nu}(t)$ represents the artificially increased measurement 
noise, assumed to be uncorrelated between ``time" samples, i.e.
\begin{equation}
E\nu(t)\nu(t')^{\rm T} = {\mathbf R}_\nu \delta(t-t')\,,
\label{eq8}
\end{equation}
where ${\mathbf R}_\nu$ is an appropriately scaled version 
of ${\mathbf C}_\nu$.

The solution procedure is obtained from a hierarchy of integro-differential 
equations involving densities of
all orders. Fortunately, the hierarchy can be truncated, to good approximation, at the first member. We then obtain
\begin{equation}
\frac{\partial\rho}{\partial t} + \phi_1\rho = 0\,,
\label{eq9}
\end{equation}
where $\phi_1$ is the conditioning factor, given by
\begin{equation}
\phi_1 = -\,({\mathbf d} - {\mathbf A\rho})^{\rm T} {\mathbf R_\nu}\!\!\!^{-1}{\mathbf A}
+ {\mathbf b}/2,
\label{eq10}
\end{equation}
and ${\mathbf b}$ represents a vector formed from the diagonal elements of ${\mathbf A}^{\rm T}{\mathbf R_\nu}\!\!\!^{-1}{\mathbf A}$. 

The optimal ${\mathbf \rho}$, denoted $\hat{\mathbf\rho}$, is obtained by 
numerically integrating 
Eq. (\ref{eq9}) from $t=0$ to $t=1$ in steps of size $\delta t$, chosen to be
small enough that the integrand changes approximately linearly between steps;  
the initial condition is $\rho_{(t=0)}=\eta$. The desired image cube of
differential column-density is
then obtained by mapping the set of $\hat{\rho}_n$
back onto the 3D grid of coordinates $(x_n,y_n,T_n)$.

We refer to the quantity $\eta$ as the {\it a priori\/} 
dilution;  it represents the degree to which the procedure is 
forced to represent the data 
with the least number of objects. In principle, $\eta$ should be set at
the smallest value for which the reduced chi squared, $\chi_\nu^2$, is
of order unity, where:
\begin{equation}
\chi_\nu^2=\frac{1}{M}\sum_i \frac{({\mathbf d}-{\mathbf A\hat{\rho}})_i^2}{(C_\mu)_{ii}},
\label{eq11}
\end{equation}
and $M$ is the number of measurements, i.e. the total number of pixels at 
all wavelengths. In practice, values in the range 0.1--0.01 typically
suffice. Provided $\eta$ (or equivalently, $N_0$) has been appropriately
chosen, the final number of representative objects 
(equal to $\sum_n{\hat{\rho}_n}$) should correspond approximately to $N_0$.

Having obtained $\hat{\mathbf\rho}$, the corresponding uncertainties 
may be obtained from the matrix of 2nd derivatives 
of $\ln P({\mathbf\rho}|{\mathbf d})$ with respect to the components of 
${\mathbf\rho}$, using the procedure described by \citet{whal71}. 
Based on the Gaussian approximation of Eq. (\ref{eq3a}), making use
of Eqs. (\ref{eq5}) and (\ref{eq6}), we construct 
a matrix, ${\mathbf\gamma}$, as follows:
\begin{equation}
{\mathbf\gamma} = {\mathbf A}^{\rm T}{\mathbf C_\mu}^{-1}{\mathbf A}
\,\,+\,\, \frac{1}{\eta}\,{\mathbf I}
\label{eq11a}
\end{equation}
where ${\mathbf I}$ is the identity matrix of order $N_{\rm st}$. 
The uncertainties in the $\hat{\rho}_n$ values are then given by:
\begin{equation}
\sigma_{\hat{\rho}_n} = [(\gamma^{-1})_{nn}]^{\frac{1}{2}}
\label{eq11b}
\end{equation}
The uncertainties become larger in localised regions of
high source density, where the number of objects required to represent
the astrophysical structure greatly exceeds the originally assumed $N_0$.
In such regions a better 
approximation is provided by replacing $\eta$ in Eq. (\ref{eq11a})
with its {\it a posteriori\/} 
value, $\hat{\eta}$, given by:
\begin{equation}
\hat{\eta} = (\sum_{n=1}^{N_{\rm st}}\hat{\rho}_n)/N_{\rm st}
\end{equation}

The above approach is motivated by two important 
considerations: \\
\indent (i) Direct maximisation of the {\it a posteriori\/} probability 
would involve searching a prohibitively large parameter space. For example, 
if we characterise each of $N_{\rm obj}$ objects by $N_{\rm p}$ parameters 
(which in this case would be $x,y,T,$ and differential column-density), we 
would need to search an $N_{\rm obj}N_{\rm p}$-dimensional space for the 
maximum probability (or, equivalently, the minimum of a weighted chi squared 
function) and this would be computationally intractible for any reasonable 
model size.  By contrast, the procedure described above is guaranteed to 
reach the globally optimal solution in a limited number of steps without 
searching a multi-object parameter space. \\
\indent (ii) The use of an occupation number formalism means that the computational burden does not increase with the number of objects.

\section[]{Tests with synthetic data}

\subsection[]{Prestellar core model}

\begin{figure}
\includegraphics[width=84mm]{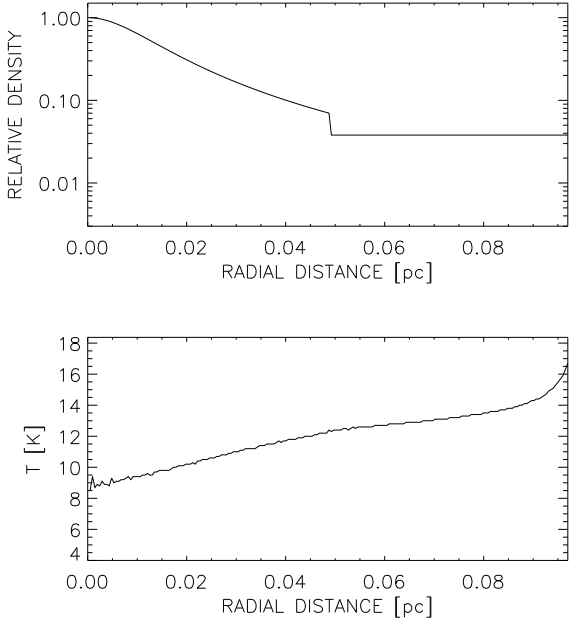}
\caption{Radial profiles in the modelled prestellar core. (a) Relative 
density, $n_{{\rm H}_2}/(1.65\times 10^5\,{\rm cm}^{-3})$. (b) Temperature, 
$T[{\rm K}]$.} 
\label{fig1}
\end{figure}

Our first test of \ppmap was based on synthetic data for a 
$0.8\,{\rm M}_{_\odot}$ 
prestellar core, modelled as a critical Bonnor-Ebert sphere with central 
density $n_{_{{\rm H}_2}}=1.65\times 10^5\,{\rm cm}^{-3}$ and radius 
$R=0.049\,{\rm pc}$, embedded in a cloud of visual absorption, 
$A_V=1\,{\rm mag}$, and located at a distance of $140\,{\rm pc}$. The radial 
profile of dust temperature for this model, and the isophotal maps observable 
with {\it Herschel\/} were computed using the {\tt PHAETHON} radiative transfer 
code \citep{stam03}. The model radial profiles of density and temperature are 
shown in Fig. \ref{fig1}.

\begin{figure}
\hspace*{1.3cm}\includegraphics[width=60mm]{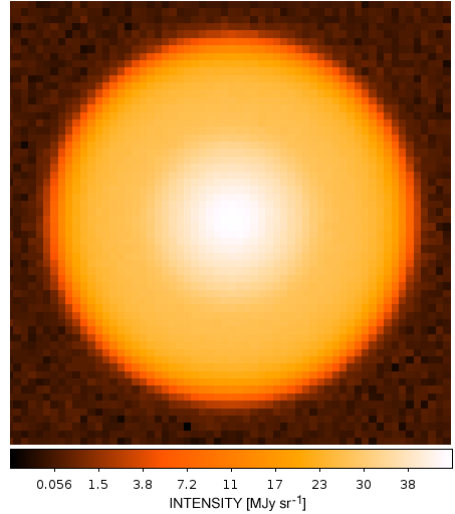}
\caption{Simulated observational image of the model prestellar core 
at $\lambda= 250\,\mu{\rm m}$. The width of the field of view is $6'$, 
corresponding to $0.24\,{\rm pc}$ at the assumed distance of $140\,{\rm pc}$.}
\label{fig2}
\end{figure}

The model profiles were used to generate synthetic {\it Herschel\/} images 
at the SPIRE/PACS nominal wavelengths of $\lambda[\mu{\rm m}]= 
70,\,160,\,250,\,350\;{\rm and}\;500$, by calculating the intensity 
distribution on the plane of the sky, using the dust opacity law 
defined by Eq. (\ref{eq12}).
These intensity distributions were then convolved with the PACS and SPIRE PSFs 
for the appropriate wavelength bands \citep{pog10,griffin13} and synthetic 
Gaussian measurement noise is added, based on an assumed SNR
of 300 at all bands. Fig. \ref{fig2} shows the result for 
$\lambda= 250\,\mu{\rm m}$.

\subsubsection[]{Results obtained using standard procedure}

\begin{figure}
\includegraphics[width=84mm]{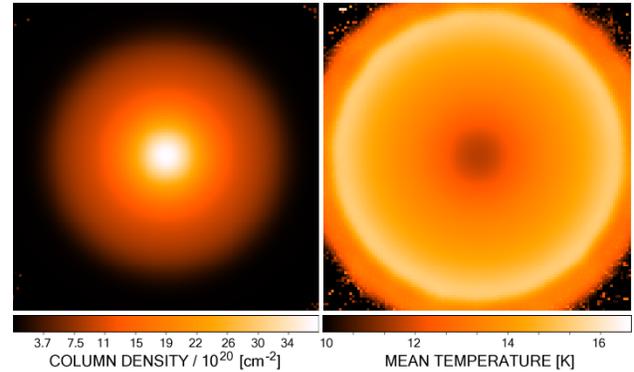}
\caption{Maps of (a) column-density, and (b) dust temperature, for the model prestellar core, obtained by applying the standard procedure, in which the dust temperature is assumed to be uniform along the
line of sight. The field of view is $6' \times 6'$.}
\label{fig3}
\end{figure}

Before processing the synthetic images with {\tt PPMAP}, we first examine the 
results obtained by applying the standard procedure. In the latter, all the 
maps are smoothed to the resolution of the $500\,\mu{\rm m}$ image, each 
pixel is then allocated a mean temperature on the basis of its SED, and 
finally this temperature is used to estimate the column-density of each pixel. 
The results are shown in Fig. \ref{fig3}.

\begin{figure*}
\includegraphics[width=160mm]{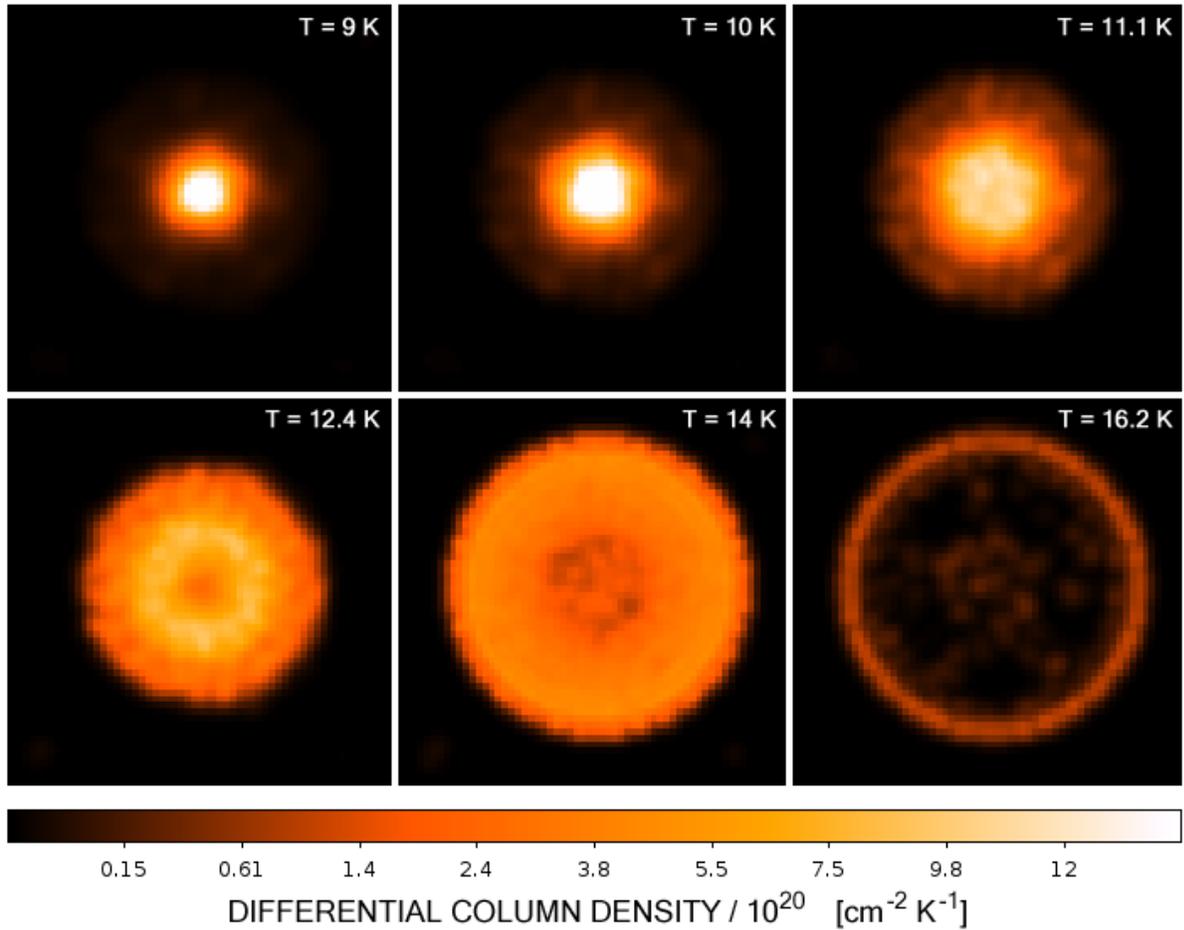}
\caption{Maps of differential column-density on six temperature planes at 
$T[{\rm K}]=9.0,\,10.0,\,11.1,\,12.4,\,14.0\;{\rm and}\;16.2$, computed 
using \ppmap on simulated observations of the model prestellar core. The 
field of view of each panel is $6' \times 6'$.}
\label{fig4}
\end{figure*}

\begin{figure*}
\includegraphics[width=160mm]{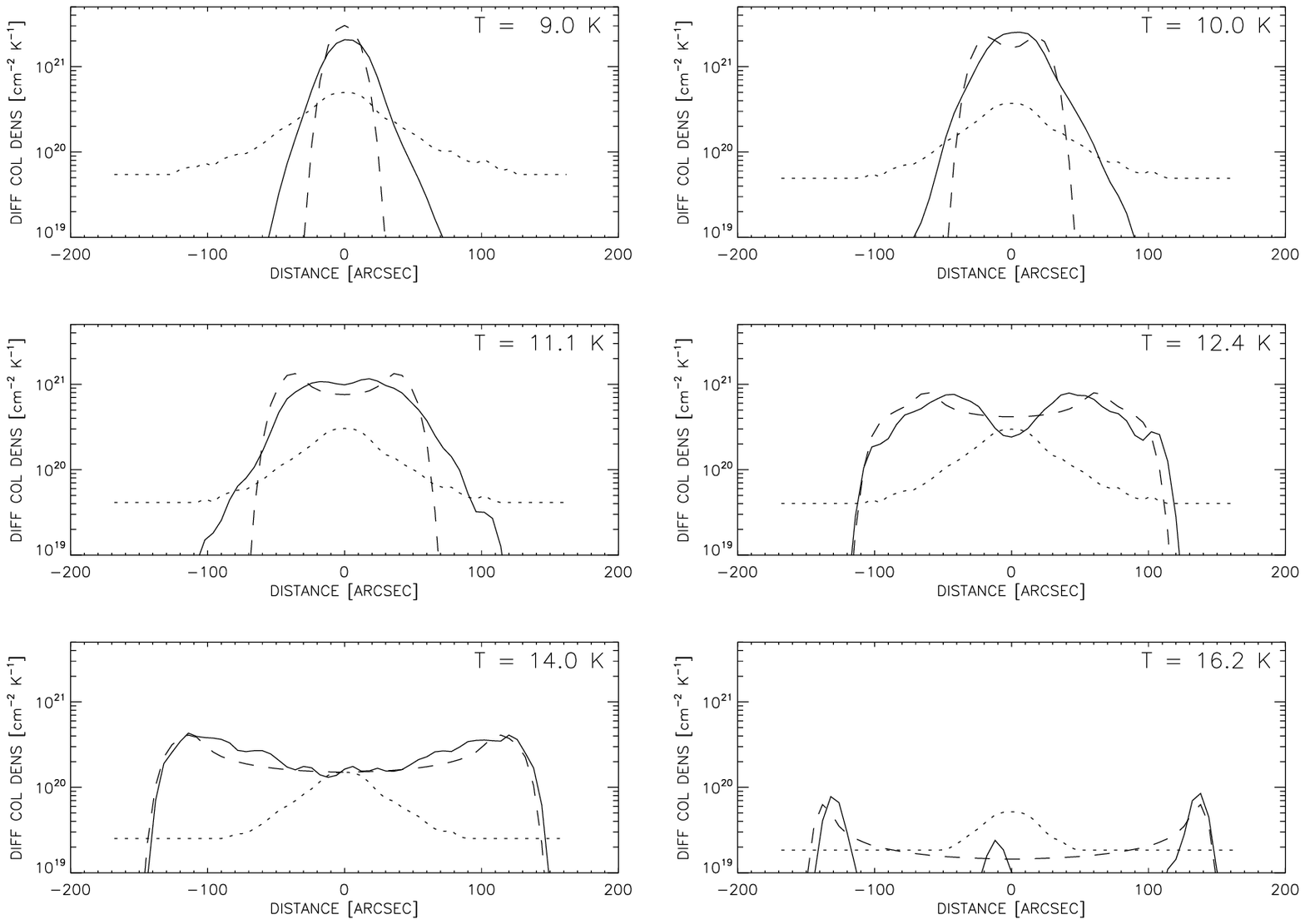}
\caption{Profiles of differential column-density on lines through the centre 
of the model prestellar core, for the temperature planes at 
$T[{\rm K}]=9.0,\,10.0,\,11.1,\,12.4,\,14.0\;{\rm and}\;16.2$. The solid lines are obtained with {\tt PPMAP}; the dashed lines are the true profiles; and the dotted lines give the 1-$\sigma$ uncertainty levels.}
\label{fig5}
\end{figure*}

\subsubsection[]{Results obtained using \ppmap}

The synthetic images were used as input to {\tt PPMAP}, to produce 
estimates of the differential column-density in a stack of ten temperature 
planes corresponding to the following set of possible dust temperatures: 
$T[{\rm K}]=7.0,\,8.0,\,9.0,\,10.0,\,11.1,\,12.4,\,14.0,\,16.2,\,19.6\;
{\rm and}\; 25.0$. The resulting differential column-density maps for the 
six temperature planes with significant values are shown in Fig. \ref{fig4}.  
They have been expressed in units of cm$^{-2}$ K$^{-1}$ by dividing the
estimated differential column-density in each temperature plane by the
temperature interval itself.
Fig. \ref{fig5} shows a central slice through these images, together with the 
true (model) profiles for comparison.

From the \ppmap results we estimate that the total mass of the model 
prestellar core plus the cloud in which it is embedded is 
$0.87\,{\rm M}_{_\odot}$, which compares well with the true mass 
of $0.85\,{\rm M}_{_\odot}$. Likewise, the \ppmap results give the 
integrated column-density along the line of sight through the centre of the 
model prestellar core to be $6.88\times 10^{21}\,{\rm H}_{_2}\,{\rm cm}^{-2}$, 
which compares well with the true value of 
$6.58\times 10^{21}\,{\rm H}_{_2}\,{\rm cm}^{-2}$. By comparison, the 
standard procedure (i.e. smoothing all images to the lowest common resolution 
and assuming constant temperature along each line of sight) gives a peak 
column-density of $3.74\times 10^{21}\,{\rm H}_{_2}\,{\rm cm}^{-2}$, i.e. 
too low by almost a factor of 2. 

\subsection[]{Fractal turbulent cloud model}

\begin{figure*}
\includegraphics[width=160mm]{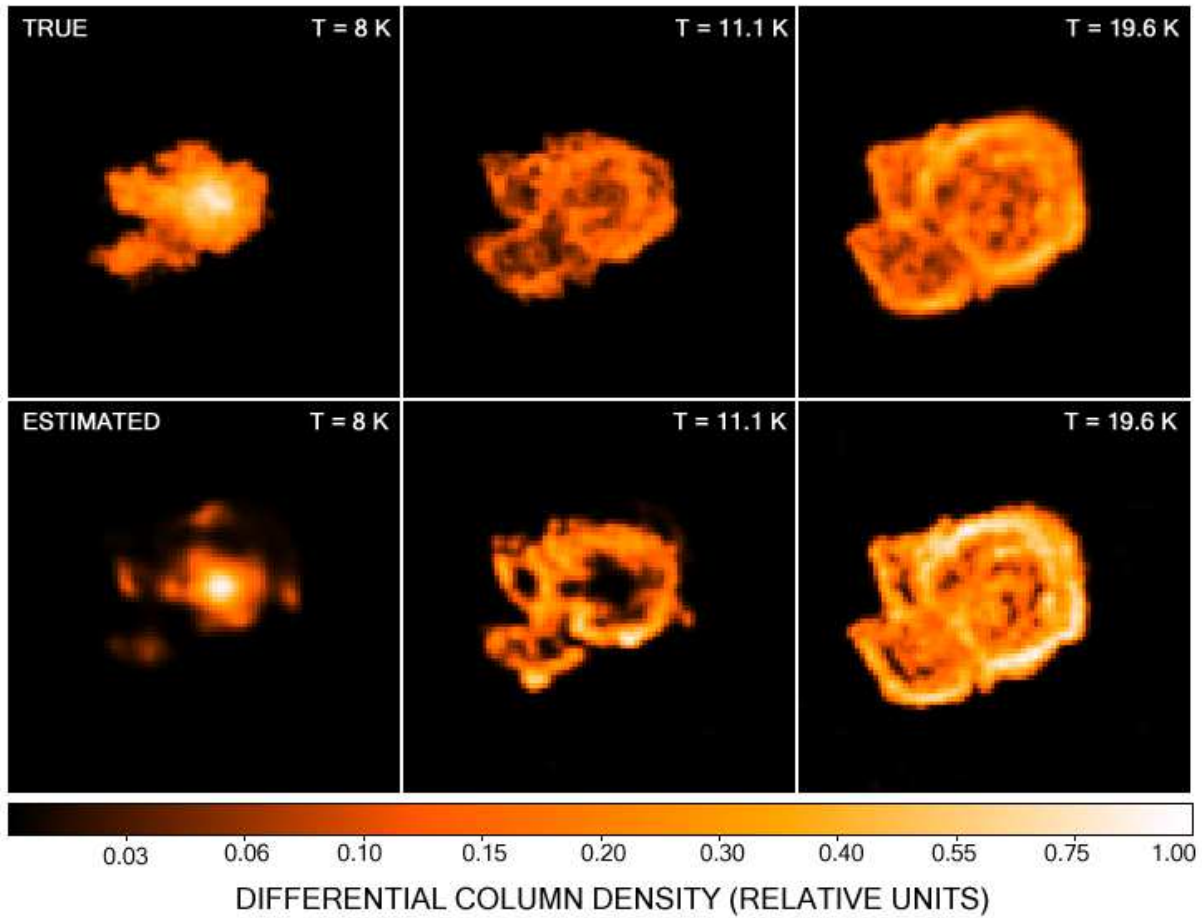}
\caption{Maps of differential column-density for the fractal turbulent
cloud model. The upper row shows the true distributions in three representative
temperature intervals centred on 8 K, 11.1 K, and 19.6 K, each with a field
of view of $9'\times9'$ for an assumed distance of 1 kpc. The lower row shows 
the inversion results obtained using {\tt PPMAP}. Within each temperature 
interval the ``true" and ``estimated" maps are presented on the same intensity
scale, but the map pairs at different temperatures have been normalised to
the same peak value of 1.0 in order to bring out the low-level structure.
The actual peak values of differential column density, in units
of $10^{22}\,{\rm cm}^{-2}{\rm K}^{-1}$, are 6.1, 2.8, and 0.15, at
the three temperatures, respectively.}
\label{fig6}
\end{figure*}

\begin{figure}
\hspace*{1.3cm}\includegraphics[width=60mm]{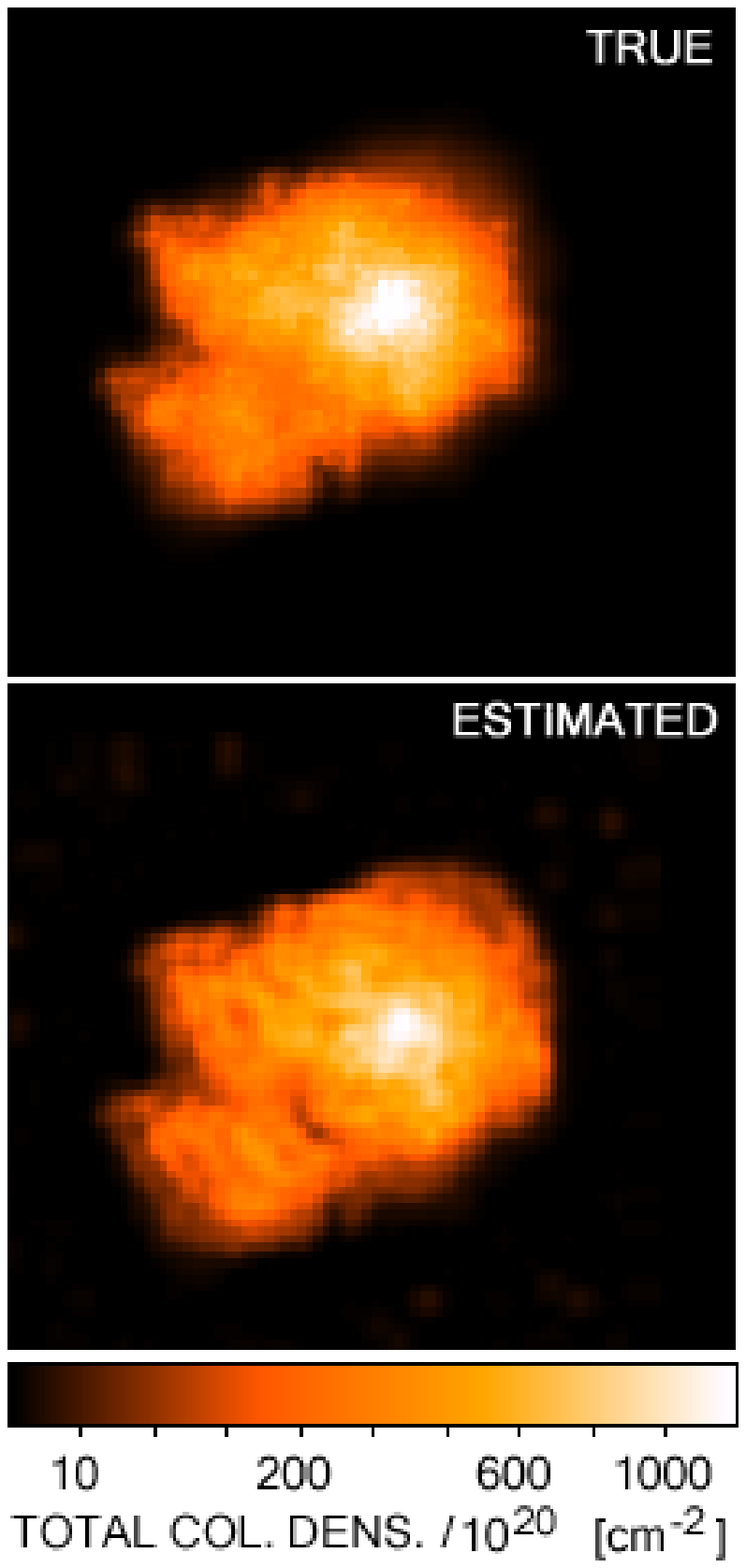}
\caption{The line-of-sight integrated column density for the fractal
turbulent cloud model, with the same field of view as for Fig. \ref{fig6}.
The top panel shows the true distribution, and the bottom panel shows the
estimated version obtained by summing the differential column density,
obtained from \ppmap, over all of the temperature intervals.}
\label{fig7}
\end{figure}

We have also tested \ppmap using synthetic data for a model turbulent cloud with
fractal structure, i.e., a nested, self-similar hierarchy of clumps
within clumps, as described by \citet{walch11}. We chose a model density 
distribution with total mass 1000 $M_\odot$ and determined its temperature 
structure via a radiative transfer calculation, assuming it to be bathed in 
an interstellar radiation field which gives rise to
dust temperatures $\sim7$ K in the centres of clumps and
$\sim20$ K at the cloud periphery.
Synthetic {\it Herschel\/} data were generated at the same 
five wavelengths as above, assuming the structure to be at a distance
of 1 kpc. As with the prestellar core model, the emergent intensity distribution
was convolved with the {\it Herschel\/} PSFs, and Gaussian noise added 
(SNR = 300). A set of 2D projections of the assumed 3D model density 
distribution is shown in the upper portions of Figs. \ref{fig6} 
and \ref{fig7}, which represent summations over selected temperature intervals
and the total line-of-sight, respectively.

Table 1 presents a summary of the results obtained in the testing of
\ppmap on synthetic data for both models, i.e., the prestellar core and
fractal turbulent cloud. Wherever possible it includes a comparison with
results obtained using the standard techniques for column density mapping.
Some conclusions which can be drawn are:
\begin{enumerate}
  \item For both the prestellar core and fractal cloud, \ppmap yielded
peak column densities, total masses and minimum dust temperatures close to the 
true values, while conventional techniques of column density mapping gave 
underestimates for peak column density and total mass, and overestimates
for the minimum temperature. 
 \item The apportioning of mass between the different temperature intervals
was more accurate in the case of the prestellar core than for the fractal 
cloud, the median values of fractional error being 15\% and 55\%, respectively.
\end{enumerate}

\begin{table*}
 \centering
 \begin{minipage}{140mm}
  \caption{Results of testing with synthetic data.}
  \begin{tabular}{@{}llcccccccccccccc@{}}
  \hline
  &  & & &  \multicolumn{4}{c}{PRESTELLAR CORE} 
& & & \multicolumn{4}{c}{FRACTAL TURBULENT CLOUD} \\
   Property & Units & & & True & \ppmap & Std.\footnote{Standard technique for
the mapping of integrated column density along the line of sight
\citep[see, for example,][]{kon10}.} & Std.(enhanced)\footnote{Enhanced version 
of standard technique whereby spatial filtering is used to improve the 
resolution \citep{palm13}.} & & & True & \ppmap & Std. & Std.(enhanced) \\
 \hline
$N({\rm H}_2)_{\rm peak}$ & [$10^{21}{\rm cm}^{-2}$] & & & 
    6.58 & 6.88 & 3.74 & 4.22 & & & 125 & 115 & 45 & 47 \\
$T_{\rm min}$\footnote{Estimated lowest temperature present in the structure
 (corresponding to the central value in the case of the prestellar core).} & 
[K]  & & & 9.0 & 9.0 & 11.9 & 11.7 & & & 7.0 & 7.0 & 11.4 & 13.2 \\
$M(T=7\,{\rm K})$\footnote{Total mass at that temperature
obtained by summing, over the angular field of view, the differential column 
density within the corresponding temperature interval.} & [$M_\odot$] & & &
    0.00 & 0.00 & -- & -- & & & 107 & 9 & -- & -- \\
$M(T=8\,{\rm K})$ & [$M_\odot$] & & & 0.00 & 0.00 & -- & -- & & & 237 & 107 & -- & -- \\
$M(T=9\,{\rm K})$ & [$M_\odot$] & & & 0.02 & 0.04 & -- & -- & & & 174 & 268 & -- & -- \\
$M(T=10\,{\rm K})$ & [$M_\odot$] & & & 0.07& 0.08 & -- & -- & & & 121 & 221 & -- & -- \\
$M(T=11.1\,{\rm K})$ & [$M_\odot$] & & & 0.13 & 0.11 & -- & -- & & & 93 & 130 & -- & -- \\
$M(T=12.4\,{\rm K})$ & [$M_\odot$] & & & 0.27 & 0.24 & -- & -- & & & 79 & 72 & -- & -- \\
$M(T=14\,{\rm K})$ & [$M_\odot$] & & & 0.31 & 0.37 & -- & -- & & & 70 & 28 & -- & -- & \\
$M(T=16.2\,{\rm K})$ & [$M_\odot$] & & & 0.04 & 0.04 & -- & -- & & & 62 & 43 & -- & -- \\
$M(T=19.6\,{\rm K})$ & [$M_\odot$] & & & 0.00 & 0.00 & -- & -- & & & 47 & 73 & -- & -- & \\
$M(T=25\,{\rm K})$ & [$M_\odot$] & & & 0.00 & 0.00 & -- & -- & & & 9 & 4 & -- & -- & \\
$M$(total) & [$M_\odot]$  & & & 0.85 & 0.87 & 0.80 & 0.81 & & & 1000 & 954 &612&612\\
\hline
\end{tabular}
\end{minipage}
\end{table*}

With regard to item (i), the superior performance of \ppmap can be
attributed to the fact that it takes full account of line-of-sight
temperature variations. Regarding (ii) it is evident that, for both models, the
mass errors for individual temperatures are
much larger than the error in total mass, and this
reflects the high degree of correlation between the errors.
In all cases the mass errors are consistent with the expected
uncertainties which, for the fractal cloud are $\sim85\,M_\odot$ at 
temperatures in the range 7--14 K, decreasing to $4M_\odot$ at 25 K.
As to the question of why the errors are significantly larger for the fractal 
cloud than for the prestellar core, the difference probably reflects the 
information content of the observations relative to the complexity of 
either model. In particular, Table 1 shows that the prestellar core
model has significant mass for only 6 temperatures,
whereas the fractal cloud model has significant values for 10 temperatures.
The 5 observational wavelengths (in conjunction with prior information) are 
apparently sufficient to constrain the 6 temperatures of the prestellar core
but insufficient to constrain the 10 temperatures of the fractal cloud.
In the latter case there is some degeneracy in the tradeoff of differential
column density between neighbouring temperatures. When understood in these
terms, the difference between the estimated and true values of fractal cloud 
mass at different temperatures is not as alarming as the 55\% error 
would suggest, since closer inspection shows that the only significant
difference is that the peak of the distribution is pushed upwards by 
$\sim1$ K. The use of more observational wavelengths would better constrain
the distribution.

\section[]{Application to real data}

We have applied \ppmap to {\it Herschel\/} data for a region of active star 
formation in the Galactic plane. The region, part of the CMa OB1 giant 
molecular cloud, was observed at wavelengths $\lambda[\mu{\rm m}]
\simeq 70,\,160,\,250,\,350\;{\rm and}\;500$, as part of the Hi-GAL survey 
\citep{mol2010}, and is described in detail by \citet{elia13}. We have 
analysed a $12'\!\!.8 \times 12'\!\!.8$ region centred on $[\ell,b]=
[224.\!\!^\circ 2717, -0.\!\!^\circ8361]$, which is dominated by a filamentary 
ridge at the western periphery of a prominent cavity.  Colour corrections 
were not applied in this inversion, i.e., we assumed $K_\lambda(T)=1$.
This was because, for the dust temperatures under consideration, the 
deviations of the correction factors from unity
are relatively small over most of the wavelength range 
\citep[$\stackrel{<}{_\sim}3$\% for  160--500 $\mu$m;][]{sad13}. Even
at 70 $\mu$m, for which the correction is larger, it is still not significant
compared to model errors since, in the field under study,
essentially all of the 70 $\mu$m emission is due to protostellar point sources 
which are not well modelled by optically thin dust. Colour corrections would, 
however, be necessary for other fields in which extended 70 $\mu$m dust 
emission is present.

\begin{figure*}
\includegraphics[width=160mm]{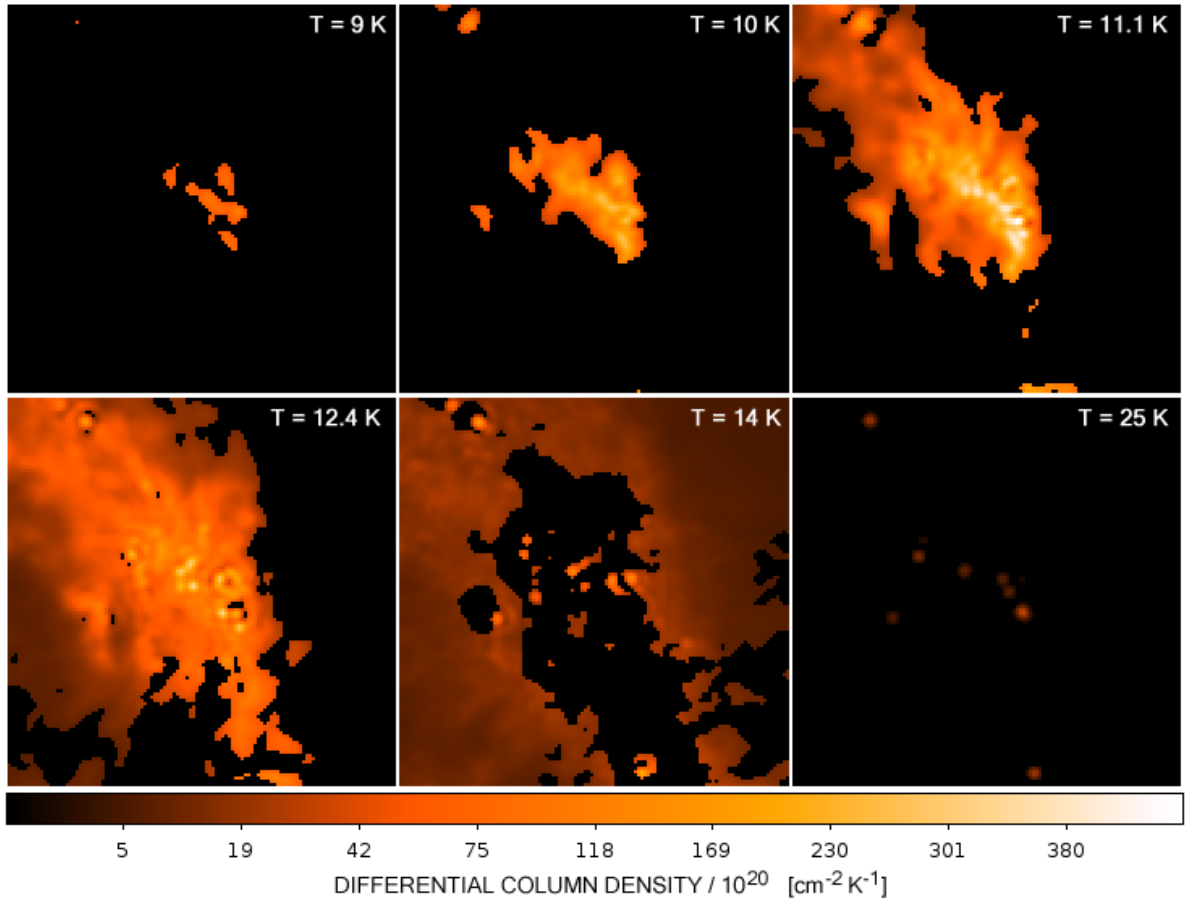}
\caption{Maps of differential column-density on six representative temperature 
planes, $T[{\rm K}] = 9.0,\,10.0,\,11.1,\,12.4,\,14.1\;{\rm and}\;25.0$, 
for a filament in the CMa OB1 molecular cloud at $\ell\simeq 224^\circ$, 
observed as part of the Hi-GAL 
survey. The display scale in each panel has been truncated at the 
corresponding 1-$\sigma$ uncertainty level.
The differential column-densities in the highest temperature panel ($T=25$ K) 
represent upper bounds when expressed per unit temperature, since the 
corresponding temperature interval has no upper bound; the displayed values 
for that particular panel are based on $\Delta T=2.7$ K.
The field of view of each panel is $12'\!\!.8\times 12'\!\!.8$.}
\label{fig8}
\end{figure*}

\begin{figure}
\includegraphics[width=84mm]{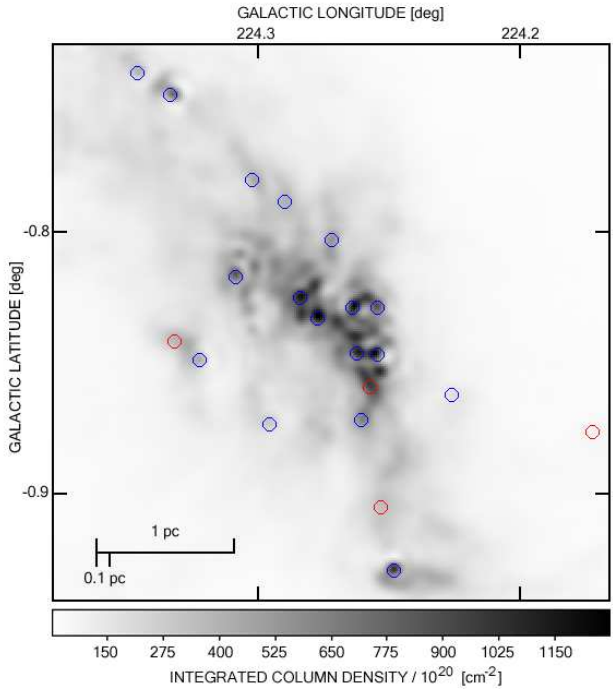}
\caption{Integrated column-density map of the filamentary complex,
obtained by summing the differential column-densities over all temperature
planes. The estimated peak column density of hydrogen molecules is 
$1.35\times10^{23}\,{\rm cm}^{-2}$. Prestellar and protostellar cores extracted by \citet{elia13}
are overplotted as red squares and blue circles, respectively. The field
of view is the same as for Fig. \ref{fig8}.}
\label{fig9}
\end{figure}

\begin{figure}
\includegraphics[width=90mm]{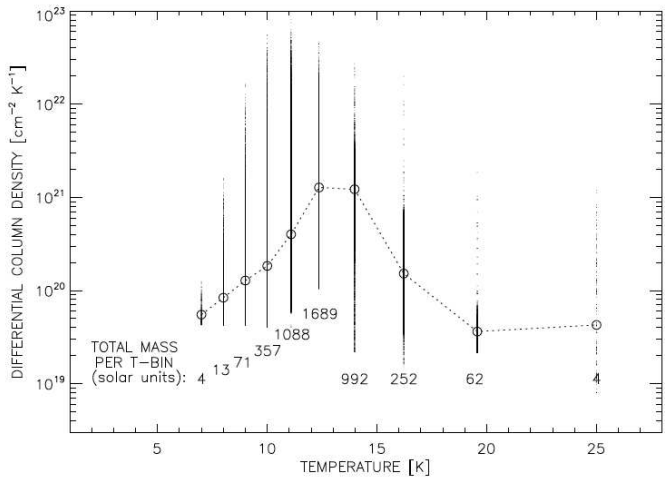}
\caption{Differential column density as a function of temperature for
the filamentary complex. At each discrete temperature, the individual 
plotted points represent
single locations in the 3D space of position and temperature. Open circles
represent the median values at each of those temperatures. Below each circle
is the total mass of material in the corresponding temperature bin.}
\label{fig10}
\end{figure}

Since the computational 
cost of \ppmap scales approximately as the square of the number of image 
pixels, we reduced the computation time by analysing the region 
as a $3\times 3$ mosaic of partially overlapping $6'\times 6'$ fields. The 
resulting maps of differential column-density are shown for six representative 
temperature ranges in Fig. \ref{fig8}; the integrated column-density map, 
obtained by summing the differential column-density over all temperatures, is 
shown in Fig. \ref{fig9}.
Fig. \ref{fig10} shows a plot of differential column density as a function
of temperature, and includes the total mass at each temperature.

Uncertainties in the differential column density estimates result from
a combination of random and systematic effects. The former are due to
measurement noise and are well represented by Eq. (\ref{eq11b}). That equation, 
however, does not take into account the systematic errors associated with
flux calibration. The correlation of those errors
between bands could, in principle, result in systematic effects in
temperature estimation. However, based on the results of
\citet{sad13} who simulated such effects
for the combination of PACS and SPIRE data, we estimate that the effect
of flux calibration errors (including the correlated component) contributes 
less than 1 K to our temperature uncertainties.

Different structures are visible in the different temperature planes in 
Fig. \ref{fig8}. These range from dense cores seen 
at $T[{\rm K}]=9\;{\rm and}\;10$, to protostars seen at $T=25\,{\rm K}$; 
at the latter temperature, the contribution of the background interstellar 
medium has disappeared. At the estimated distance of 
$1.1\,{\rm kpc}$ \citep{elia13}, the cores are only marginally resolved and 
therefore they do not show the shell structure evident in our maps of the 
model prestellar core (Fig. \ref{fig4}). A shell structure is, however,
evident in the filamentary envelope which shows a characteristic depression at
$T=14$ K indicative of the lack of interior warm material.

The total mass of the filamentary complex within the 
analysed $12'\!\!.8\times12'\!\!.8$ region, obtained by summing 
contributions from the pixels of the integrated column-density map of 
Fig. \ref{fig9}, is $M_{_{\rm TOT}}\simeq 4500\,{\rm M}_{_\odot}$. The 
mean mass per unit length of the {\it large structure} (i.e. the structure 
filling the frame of Fig. \ref{fig9}) is 
then $\mu_{_{\rm FIL}}\sim 1000\,{\rm M}_{_\odot}\,{\rm pc}^{-1}$. This is 
about two orders of magnitude larger than the maximum equilibrium value 
for an isothermal cylinder at $10\,{\rm K}$, i.e. $\sim\!16\,{\rm M}_{_\odot}\,
{\rm pc}^{-1}$ \citep{inut97}, which is consistent with the highly 
fragmented appearance of the large structure.  Many local maxima are
visible and the most prominent of these correspond to compact sources
extracted by \citet{elia13}; the locations of the extracted prestellar
and protostellar cores are overplotted on Fig. \ref{fig9}.

\begin{figure}
\includegraphics[width=90mm]{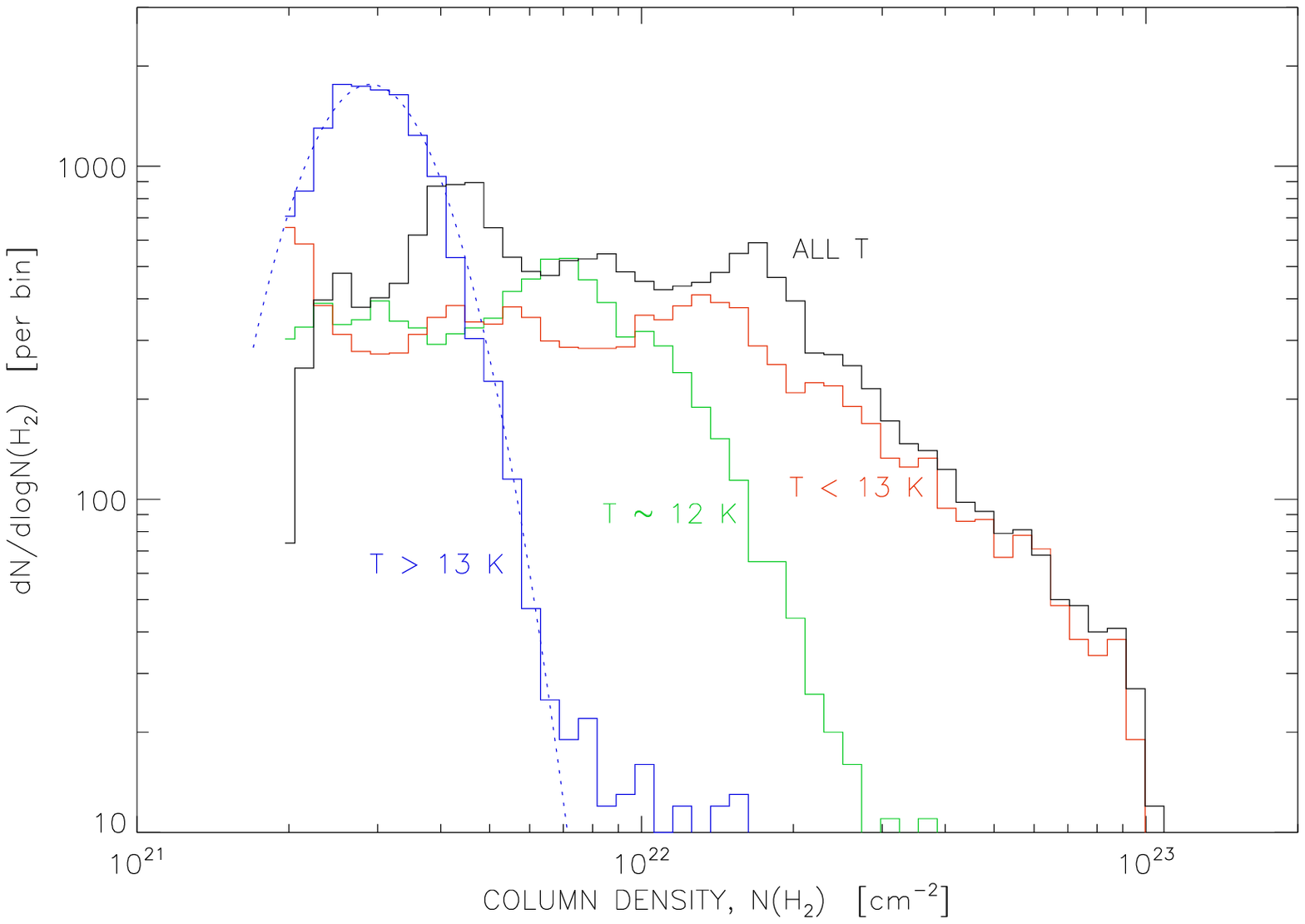}
\caption{The distribution of column-densities in the filamentary complex.
The black histogram represents the distribution of integrated column-density.
The red, green, and blue
histograms represent the column-densities of material with
$T<13$ K, $T\sim12$ K, and $T>13$ K, respectively.  The blue dotted line 
represents a log-normal function with standard deviation of log 
column-density equal to 0.28.}
\label{fig11}
\end{figure}

The distribution of dust-derived column-density values for the larger 
($\sim9^\circ\times 2^\circ$) field, which includes the filamentary complex, 
is presented as a histogram by \citet{elia13}; it can be characterised
as a log-normal distribution with a power-law tail, consistent with
the effects of self-gravity on density fluctuations produced by interstellar
turbulence \citep{klessen01,kain09,krit11,sch13}. The column-density
distribution within the $12'\!\!.8\times12'\!\!.8$ region under
present study is shown by the black histogram in Fig. \ref{fig11}. In
contrast to the \citet{elia13} plot, no log-normal component is apparent;
the most prominent feature is a power-law-like variation at high column
densities.  
However, if instead of considering total column-density we sum the 
{\it differential\/} column-density over various separate temperature ranges, 
we obtain a somewhat different picture. This is evident, for example, from the 
blue and red histograms in Fig. \ref{fig11}, which represent the column-density 
distributions of the warm ($T>13$ K) and cool ($T<13$ K) material, respectively.
Of these, the blue histogram is well fit by a log-normal, whose 
peak location ($2.9\times10^{21}\,{\rm cm}^{-2}$) and standard deviation
of log column-density (0.28) closely match the log-normal component plotted by 
\citet{elia13}---{\it the latter component, therefore, is still present\/},
even though not apparent in the histogram of integrated column-density
for the $12'\!\!.8\times12'\!\!.8$ field. The green histogram represents
material at $T\sim12$ K. It appears that the material at this intermediate
temperature dominates the flat portion of the total histogram below a
column density of $\sim2\times10^{22}\,{\rm cm}^{-2}$. It is also
the temperature range in which the differential column density reaches its
peak, as shown by Fig. \ref{fig10}.

These results might be interpreted to mean that the warm gas
(log-normally distributed) has retained the
density structure produced by interstellar turbulence while the cool
gas has collapsed into cores and comprises the high density tail of the
histogram.  The reason that the log-normal (turbulent)
component is much more prominent in the \citet{elia13} histogram is that
the latter was derived from a much larger area of sky, over which the total
contribution of the warm ISM component was significantly
larger than that of the more localised filamentary structure.  
The material at intermediate temperatures ($\sim12$ K) may 
be in a transitional stage of evolution, whereby self gravity has
taken over, but the collapse has not yet terminated in a power-law
distribution. Such a scenario is consistent with the simulations of
\citet{ward14}. 

We defer a more quantitative analysis of the distribution of
differential column density to a forthcoming paper. In that regard
we expect that the decomposition of column densities into components at 
different temperatures will help in resolving some of the issues,
currently being debated, in the
interpretation of column density PDFs. These include the question of
whether the apparent power-law tail can be interpreted more fundamentally
as a combination of log-normals \citep{brunt15}, or whether the reverse
is true, i.e. that the apparent log-normal components are actually
combinations of power-laws with low-column-density turnovers 
\citep{lom15}.

\section[]{Discussion}

The \ppmap results demonstrate that there is considerably 
more information in multi-wavelength imaging data than simply the integrated 
column-density and the mean dust temperature.  Even if only the integrated 
column-density is required, \ppmap provides more accurate estimates 
of both peak column density and total mass than the standard analysis 
procedure.  The increased accuracy 
derives both from the ability to capture line-of-sight temperature variations, 
and from the improved spatial resolution that comes with not having to smooth 
observational data to the lowest common resolution. The minimum temperature
along the line of sight can also be obtained with much greater accuracy,
and this is particularly important in the study of starless cores whereby
the gas chemistry at core centre is strongly temperature dependent.
Moreover, {\tt PPMAP}-based estimates of column-density and temperature at 
the centre of a starless core are model- and geometry-{\it independent\/}. 

The multi-temperature maps of differential column-density can 
aid in the interpretation of more complex systems by distinguishing
different physical components along the line of sight, as illustrated
by our analysis of the filamentary structure in the CMa OB1 cloud.
In particular the image cubes served to bring out the log-normal component 
which was not at all apparent in the
distribution of integrated column-density in the immediate
vicinity of the filamentary complex. Our future work will include a more
quantitative analysis of the functional forms of the PDFs of differential
column density at different temperatures. In particular we expect that the 
temperature decomposition will provide some insight into the issues currently
being debated in connection with PDFs of molecular clouds.
An additional avenue that we will pursue is to add an additional variable
to our state space, namely the index, $\beta$, of the dust opacity law
in order to provide information on the spatial variation of grain properties.
This will necessitate the inclusion of submillimetre data at longer
wavelengths in order to break the well-known degeneracy between temperature and
opacity.

Although we have applied the Point Process algorithm to the problem of
column density mapping of dusty Galactic structures, the technique
itself is far more generic, and can be applied to any system that can
be represented as a set of points in a suitably defined parameter space, such 
that the instrumental response to each point contributes independently to 
the observations, i.e., the measurement model obeys the superposition 
principle. In addition the formalism is ideally suited to the study
of dynamically evolving systems. To deal with the latter, it is necessary
only to add a dynamic term of the form $L\rho$ to Eq. (\ref{eq9}), where
$L$ is the Fokker-Planck operator \citep{rich92}. This would, for example,
provide the ability to integrate on a moving object without knowing, 
in advance, how fast it is moving or in what direction. The single-object
state space would then include not only the $(x,y)$ source position, but two 
additional variables, $v_x$ and $v_y$, representing the components of source 
velocity. One important application might be the detection of near-Earth
asteroids, whereby \ppmap has the potential to provide a significant
increase in sensitivity.

\section[]{Conclusions}

\ppmap is an algorithm designed to produce image cubes of differential
column density as a function of angular position and dust temperature for
dusty astrophysical structures associated with star formation. The input
data consist of a set of observational images at various wavelengths and the
associated PSFs. All observational images are used at their native resolution 
and no smoothing is required. 

The performance has been evaluated using 
simulated {\it Herschel\/} data at five wavelengths between 70 $\mu$m and 
500 $\mu$m. Two representative cases were chosen, namely a model prestellar 
core (embedded Bonnor-Ebert sphere) and a spatially complex model of a
fractal turbulent cloud, the dust temperatures being based on a radiative
transfer model. In both cases the spatial structure at different
temperatures was recovered well. The apportioning of mass between different
temperatures was accompished accurately for the prestellar core and
reasonably well for the fractal cloud, except for a displacement in
the distribution of estimated differential column density 
by $\sim1$ K in the latter case. The displacement reflects a limitation
in the number of temperatures which can be constrained using observational
data at five wavelengths.  The temperature resolution can be expected 
improve with the use of additional observational wavelengths. Comparison
with column density maps produced by conventional techniques shows that
\ppmap can produce significantly more accurate estimates of peak column
density, total mass, and minimum dust temperature within the particular
structure.

Application of \ppmap to a filamentary complex observed during the
Hi-GAL survey shows that the decomposition into different temperatures
facilitates the separation of different physical components along the
line of sight and has the potential to provide insight into the 
mechanisms associated with column density PDFs of molecular clouds.

\section*{Acknowledgments}

We dedicate this paper to the memory of John M. Richardson, a dear friend
and former colleague of one of us (KAM), whose earlier development of 
Point Process algorithms provided the mathematical foundation of this work.
We also thank the referee for helpful comments.
This research is supported by the EU-funded {\sc vialactea} Network (Ref. FP7-SPACE-607380).

\bsp

\label{lastpage}

\end{document}